\documentstyle[eqsecnum,aps,epsf]{revtex} 
\newcommand{\postscript}[2] {\setlength{\epsfxsize}{#2\hsize}
\centerline{\epsfbox{#1}}}

\begin{document}
\twocolumn[\hsize\textwidth\columnwidth\hsize\csname@twocolumnfalse\endcsname

\title{\bf Two phase transitions in  ${ (d_{x^2-y^2}+is)}$-wave
superconductors}

\author{Angsula Ghosh and Sadhan K. Adhikari}
\address{Instituto de F\'{\i}sica Te\'orica, 
Universidade Estadual Paulista,\\
01.405-900 S\~ao Paulo, S\~ao Paulo, Brazil\\}

\date{\today}
\maketitle

\begin{abstract}

We study numerically the temperature dependencies of specific heat,
susceptibility, penetration depth, and thermal conductivity of a coupled $
(d_{x^2-y^2}+is)$-wave Bardeen-Cooper-Schrieffer superconductor in the
presence of a weak $s$-wave component (1) on square lattice and (2) on a
lattice with orthorhombic distorsion. As the temperature is lowered past
the critical temperature $T_c$, a less ordered superconducting phase is
created in $ d_{x^2-y^2}$ wave, which changes to a more ordered phase in $
(d_{x^2-y^2}+is)$ wave at $T_{c1}$. This manifests in two second-order
phase transitions.  The two phase transitions are identified by two jumps
in specific heat at $T_c$ and $T_{c1}$.  The temperature dependencies of
the superconducting observables exhibit a change from power-law to
exponential behavior as temperature is lowered below $T_{c1}$ and confirm
the new phase transition. 

{PACS number(s): 74.20.Fg, 74.62.-c, 74.25.Bt}

Keywords: $d_{x^2-y^2} +is$-wave superconductor, specific heat,
susceptibility, thermal conductivity.

\end{abstract} 

\vskip1.5pc]


The unconventional high-$T_c$ superconductors \cite{hi} with a high critical
temperature $T_c$ have  a complicated lattice structure with extended and/or
mixed symmetry for the order parameter \cite{n1,n2}.  It is generally
accepted
that for many of these high-$T_c$ materials, the order parameter exhibits 
anisotropic behavior.  However, it is difficult to establish the detailed 
nature of anisotropy, which could be of an extended $s$-wave, a pure
$d$-wave, or a mixed $(s+\exp(i\theta)d)$-wave type.  Some of the high-$T_c$
materials have singlet $d$-wave Cooper pairs  and the order parameter has
$d_{x^2-y^2}$ symmetry in two dimensions \cite{n1}.  Recent measurements
\cite{h} of the penetration depth $\lambda(T)$   and superconducting specific
heat at different temperatures $T$  and related theoretical analysis
\cite{t1,c} also support this point of view.  In some cases there is  the
signature  of an extended $s$- or $d$-wave symmetry \cite{n1,n2}. 

The possibility of a
mixed $(s-d)$-wave symmetry was suggested sometime ago by Ruckenstein et al.
and Kotliar \cite{6}. Several different types of measurements which are
sensitive to the phase of the order parameter indicate a significant
mixing of $s$-wave component  with a predominant $d_{x^2-y^2}$ state
at lower temperatures below $T_{c1}$. For temperatures between $T_{c1}$
and  
the critical
temperature $T_c$ only the $d_{x^2-y^2}$ state survives. 
 There are experimental evidences based on
Josephson supercurrent for tunneling between a conventional $s$-wave
superconductor 
(Pb) and single crystals of YBa$_2$Cu$_3$O$_7$ (YBCO)
 that  YBCO has mixed $d\pm s$ or $d\pm is$ symmetry
\cite{5} at lower temperatures. 
A similar conclusion may also be obtained based on the results
of angle-resolved 
 photoemission spectroscopy experiment by Ma et al. in which a temperature
dependent gap anisotropy 
in  Bi$_2$Sr$_2$CaCu$_2$O$_{8+x}$  was detected
\cite{7}. The measured gaps along both high-symmetry directions are
non-zero at low temperatures and their ratio was strongly temperature
dependent. This observation is 
  also difficult to reconcile employing a pure $s$-
or $d$-wave order parameter  and suggests that a mixed
[$d_{x^2-y^2}+\exp(i\theta)s$] symmetry is applicable at low temperatures.
However,
at higher temperatures one could have a pure $d_{x^2-y^2}$ symmetry of the
order parameter.
 Recently, this
idea has been explored to explain the NMR data in the superconductor YBCO and
the Josephson critical current observed in YBCO-SNS and YBCO-Pb junctions
\cite{8}. Recently, a new class of c-axis Josephson tunneling experiments
are reported by Kouznetsov et al. \cite{K} in which a conventional
superconductor (Pb) was deposited across a single twin boundary of a YBCO
 crystal. In that case measurements of critical current as a
function of the magnitude and angle of a magnetic field applied in the
plane of the junction provides a direct evidence of an order parameter of  
mixed  [$d_{x^2-y^2}+\exp(i\theta)s$] symmetry in YBCO. The microwave
complex conductivity measurement in the superconducting state of high
quality YBa$_2$Cu$_3$O$_{7-\delta}$ single crystals measured at 10 GHz
using a high-Q Nb cavity also strongly suggests a multicomponent
superconducting order parameter in YBCO \cite{S}.  

More recently, Krishana et al. \cite{Kr} reported a phase transition in
the high-$T_c$ superconductor Bi$_2$Sr$_2$CaCu$_2$O$_8$ induced by a
magnetic field from a study of the thermal conductivity as a function of
temperature and applied field.  Possible interpretation of this
measurement  could be the induction
of a minor $s$ or $d_{xy}$ component with a $d_{x^2-y^2}$ symmetry with
the application of a weak field \cite{L,F}.

There have also been certain recent theoretical studies using mixed $s$- and
$d$-wave symmetries \cite{9a,9b} and it was noted that it is more likely to
realize a stable mixed $d+is$ state than a $d+s$ state considering
different
couplings and lattice symmetries. As noted by Liu et al. \cite{9a},
a stable $d+s$ solution can not be realized
on square lattice. However, in the presence of orthorhombic distortion, such a
solution can be obtained.  The special interest in these investigations was
on the temperature dependence of the order parameter of the mixed $d+is$
state within the Bardeen-Cooper-Schrieffer
(BCS) model \cite{e,t}. The study by Liu et al. \cite{9a}
of the order parameter on mixed $d+is$
state also explored the effect of  a van Hove singularity in the density
of states on the solutions of the BCS equation. 
Laughlin has provided a theoretical explanation of the observation by
Krishana et al. \cite{Kr} that at low temperatures and for weak magnetic
field a time-reversal symmetry breaking state of mixed symmetry is induced
in Bi$_2$Sr$_2$CaCu$_2$O$_8$. From a study of vortex in a $d$-wave
superconductor using a self-consistent Bogoliubov-de Gennes formalism,
Franz
and Te\'sanovi\'c \cite{F} also predicted the possibility of the creation
of a superconducting state of mixed symmetry. Although, the creation of
the mixed state in this case is speculative, they conclude that a dramatic
change should be observed in the superconducting observables when a pure 
superconducting $d_{x^2-y^2}$  state undergoes a phase transition to a
mixed-symmetry state. In many cases there are evidences of a $s$-wave
mixture with a $d_{x^2-y^2}$ state.  In view of this, in this study we 
 present an investigation of this phase transition on specific heat,
spin susceptibility,  penetration depth, and thermal conductivity.

There is no suitable microscopic theory for high-$T_c$
superconductors. Although it is accepted that Cooper pairs are formed in
such materials,
there is controversy about a proper  description of the 
normal state and 
the pairing mechanism for such 
  materials \cite{n1}.  In the absence of a suitable microscopic theory,
a phenomenological tight-binding model 
in two dimensions which
incorporate the proper
lattice symmetry within the BCS formalism \cite{n2} has been suggested
\cite{P}. This model has been
very
successful in describing many properties  of  high-$T_c$ materials
and is often used in the study of high-$T_c$ compounds. We shall use this
model in the present investigation.

 In all previous theoretical 
 studies on mixed-symmetry superconductors 
a general behavior of the temperature dependence of the order
paramers emerged, independent of lattice symmetry employed \cite{9a,9b}.  
 The tight-binding BCS model
for a mixed $d+is$ state becomes  a coupled set of equations in the two
partial waves.  The ratio of the strengths of the  $s$- and $d$-wave
interactions should lie in a narrow region in order to have a coexisting $s$
and $d$ wave phases in the case of  $d+is$ symmetry.  As the  $s$-wave
($d$-wave) interaction becomes stronger, the $d$-wave ($s$-wave) component of
the order parameter  quickly reduces and disappears and a pure $s$-wave
($d$-wave) state emerges.

In this work we study the temperature dependencies
of different observables, such as specific heat, susceptibility, penetration
depth, and  thermal conductivity, of a $ (d_{x^2-y^2}+is)$-wave BCS
superconductor with
a weak $s$-wave admixture both on square lattice and on a lattice with 
orthorhombic
distortion and find that it exhibits interesting properties. 
A pure $s$-wave
($d$-wave) superconducting observable exhibits an exponential (power-law)
dependence on temperature. 
In the case of mixed $ (d_{x^2-y^2}+is)$-wave symmetry, we
find that it is possible to have a crossover from an exponential to power-law
dependence on temperature below the superconducting critical temperature 
and a second second-order phase transition.

For a weaker $s$-wave admixture,  in the present study we
establish   in the two-dimensional 
tight-binding model (1) on square lattice and (2) on a lattice with 
orthorhombic distortion
another second-order phase transition at $T=T_{c1}<T_c$, where the
superconducting phase changes from a pure $d$-wave state for $T>T_{c1}$ to a
mixed $(d+is)$-wave  state for $T<T_{c1}$.  The specific heat exhibits two
jumps at the transition points $T=T_{c1}$ and $T=T_c$.  
The temperature
dependencies of the superconducting specific heat, susceptibility,
penetration depth and thermal conductivity change drastically at $T=T_{c1}$
from power-law behavior (typical to $d$ state with node(s) in the order
parameter on the Fermi surface)  for $T>T_{c1}$ to exponential behavior
(typical to $s$ state with no nodes) for $T<T_{c1}$. The order parameter for
the present ($d+is$) wave  does not have a node on the Fermi surface for
$T<T_{c1}$ and it behaves like a modified $s$-wave one. The
observables for the normal state are closer to the superconducting $l = 2$
state than to those for the superconducting $l=0$ state \cite{c}.
Consequently, superconductivity in $s$ wave is more pronounced than in $d$
wave.  Hence as temperature decreases the system passes from the normal state
to a ``less" superconducting $d$-wave state  at $T=T_c$ and then to a ``more"
superconducting  state with dominating $s$-wave behavior
at $T=T_{c1}$ signaling a second
phase transition. 

The pronounced change in the nature of the superconducting
state at $T=T_{c1}$ becomes very apparent from a study of the entropy. At a 
particular temperature the entropy for the normal state is larger than that
for all superconducting states signalling an increase in order in the
superconducting state. In the case of the present $ (d_{x^2-y^2}+is)$ 
state we find that
as the temperature is lowered past $T_{c1}$, the entropy of the
superconducting $ (d_{x^2-y^2}+is)$ state decreases very rapidly (not
shown explicitly
in this work) indicating  the appearance of a more ordered superconducting
phase and a second phase transition.

We base the present study on the two-dimensional  tight binding model 
which we describe below. This model is sufficiently general for considering 
mixed angular momentum states, with or without orthorhombic distortion, 
employing nearest and second-nearest-neighbour hopping integrals.  
The effective interaction is taken
to possess  an
on-site repulsion $(v_r)$ and a nearest-neighbour attraction $(v_a)$ and 
can be represented as
\begin{equation}\label{1}
V_{{\bf k}{\bf q}}=v_r-v_a[\cos (k_x-q_x)+\beta^2 \cos (k_y-q_y)],
\end{equation}
where $\beta =1$ corresponds to a square lattice, and $\beta \ne 1$
represents orthorhombic distortion. On expansion,  and
keeping only the $s$- and $d_{x^2-y^2}$-wave components of this
interaction we have    
\begin{equation}\label{2}
V_{{\bf k}{\bf q}}=-V_0-V_2(\cos k_x-\beta \cos k_y)(\cos q_x
-\beta\cos q_y).
\end{equation}
 Here $V_0=-v_r+(1+\beta^2)v_a/2$ and $V_2=v_a/2$ are the 
 couplings of effective 
 $s$- and 
 $d$-wave interactions, respectively. As we shall consider Cooper pairing and
subsequent BCS condensation in both $s$ and $d$ waves  
the constants $V_0$ and $V_2$
will be taken to be  positive  corresponding to attractive interactions. 
In this case the quasiparticle dispersion relation is given by
\begin{equation}\label{3}
\epsilon_{\bf k}=-2t[\cos k_x+\beta \cos k_y-\gamma\cos k_x
\cos k_y],
\end{equation}
where $t$ and $\beta t$ are the nearest-neighbour hopping integrals
along the in-plane $a$ and $b$ axes, respectively, and $\gamma t/2$ is the 
second-nearest-neighbour hopping integral. 
The energy $\epsilon_{\bf k}$ is  measured with respect
to the surface of the Fermi sea.

We consider the weak-coupling BCS model  in two dimensions with
$ (d_{x^2-y^2}+is)$ symmetry.  At a finite $T$, one has  the following BCS
equation
\begin{eqnarray}
\Delta_{\bf k}& =& -\sum_{\bf q} V_{\bf kq}\frac{\Delta_{ \bf q}}{2E_{\bf
q}}\tanh
\frac{E_{\bf q} }{2T}  \label{130} \end{eqnarray} with $E_{\bf q} =
[\epsilon_{\bf q} ^2 + |\Delta_{\bf q}|^{2}]^ {1/2}$. We use units
$k_B=1$, where $k_B$ is the Boltzmann constant.
The order parameter $\Delta _{\bf q}$ has the
following anisotropic form: $\Delta _{\bf q} \equiv \Delta_0+i \Delta_2 (\cos
q_x -\beta \cos q_y)$.    
Using the above form of $\Delta_{\bf q}$ and potential (\ref{2}),
Eq. (\ref{130}) becomes the following coupled set of BCS equations
\begin{eqnarray}
\Delta_0=V_0\sum_{\bf q}\frac{\Delta_0}{2E_{\bf q}}\tanh
\frac{E_{\bf q}}{2T}\label{131}
\end{eqnarray}
\begin{eqnarray}
\Delta_2=V_2\sum_{\bf q}\frac{\Delta_2(\cos k_x-\beta
\cos k_y)^2}
{2E_{\bf q}}\tanh
\frac{E_{\bf q}}{2T}\label{132}
\end{eqnarray}
where the coupling is introduced through $E_{\bf q}$. In Eqs. (\ref{131})
and (\ref{132}) both the interactions $V_0$ and $V_2$ are assumed to be 
energy-independent constants for $|\epsilon_{\bf q} | <  T_D$
and zero for $|\epsilon_{\bf q} | >  T_D$, where $ T_D$ is a mathematical 
cutoff to ensure convergence of the integrals. It should be compared with 
the  
usual Debye cutoff for the conventional superconductors.

The  specific heat per particle  is given by \cite{t}
\begin{equation}
C(T)= \frac{2}{NT^2}\sum_{\bf q}   f_{\bf q}(1-f_{\bf q})
\left( E_{\bf q}^2-\frac{1}{2}T\frac{d|\Delta_{\bf q}|^2}{dT} \right)
\label{sp} 
\end{equation} 
where $f_{\bf q}=1/(1+\exp( E_{\bf q}/ T))$.  The spin-susceptibility $\chi$
is defined by
\cite{c}
\begin{equation}
\chi(T)= \frac{2\mu_N^2}{T}\sum_{\bf q}f_{\bf q}(1-f_{\bf q})
\end{equation}
where $\mu_N$ is the  nuclear magneton.  The penetration depth $\lambda$ is
defined by \cite{t}
\begin{equation}
\lambda^{-2}(T) = \lambda^{-2}(0)\left[1- \frac{2}{NT}
\sum_{\bf q} f_{\bf q}(1-f_{\bf q})\right].
\end{equation}  
The superconducting to normal thermal conductivity ratio $K_s(T)/K_n(T)$ is
defined by \cite{c}
\begin{equation}
\frac {K_s(T)}{K_n(T)} = \frac {\sum_{\bf q} (\epsilon_q -1)f_{\bf q}(1-f_
{\bf q})E_{\bf q} }{\sum_{\bf q} (\epsilon_q -1)^2 f_{\bf q}(1-f_{\bf q})}.
\end{equation}

\vskip -2.5cm
\postscript{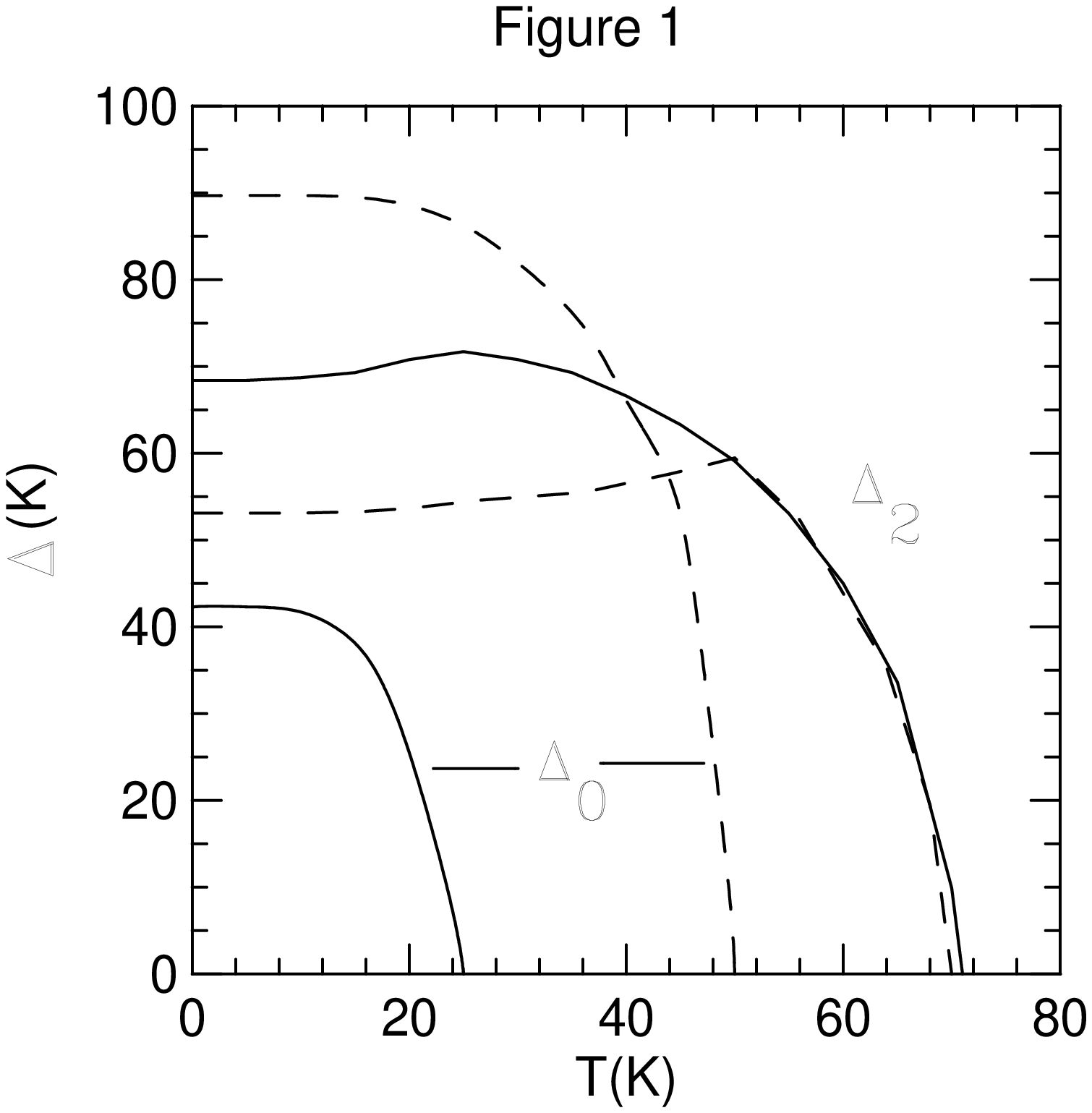}{1.0}    
\vskip -2.1cm

{ {\bf Fig. 1.}  The  $s$- and $d$-wave parameters
$\Delta_0$,  $\Delta_2$ in Kelvin 
 at different temperatures for $ (d_{x^2-y^2}+is)$-wave models 
1(a) (square lattice: full line) and 2(b) (orthorhombic distortion: 
dashed line)
described in the text with different mixtures of $s$ and $d$ waves.}

\vskip -2.5cm
\postscript{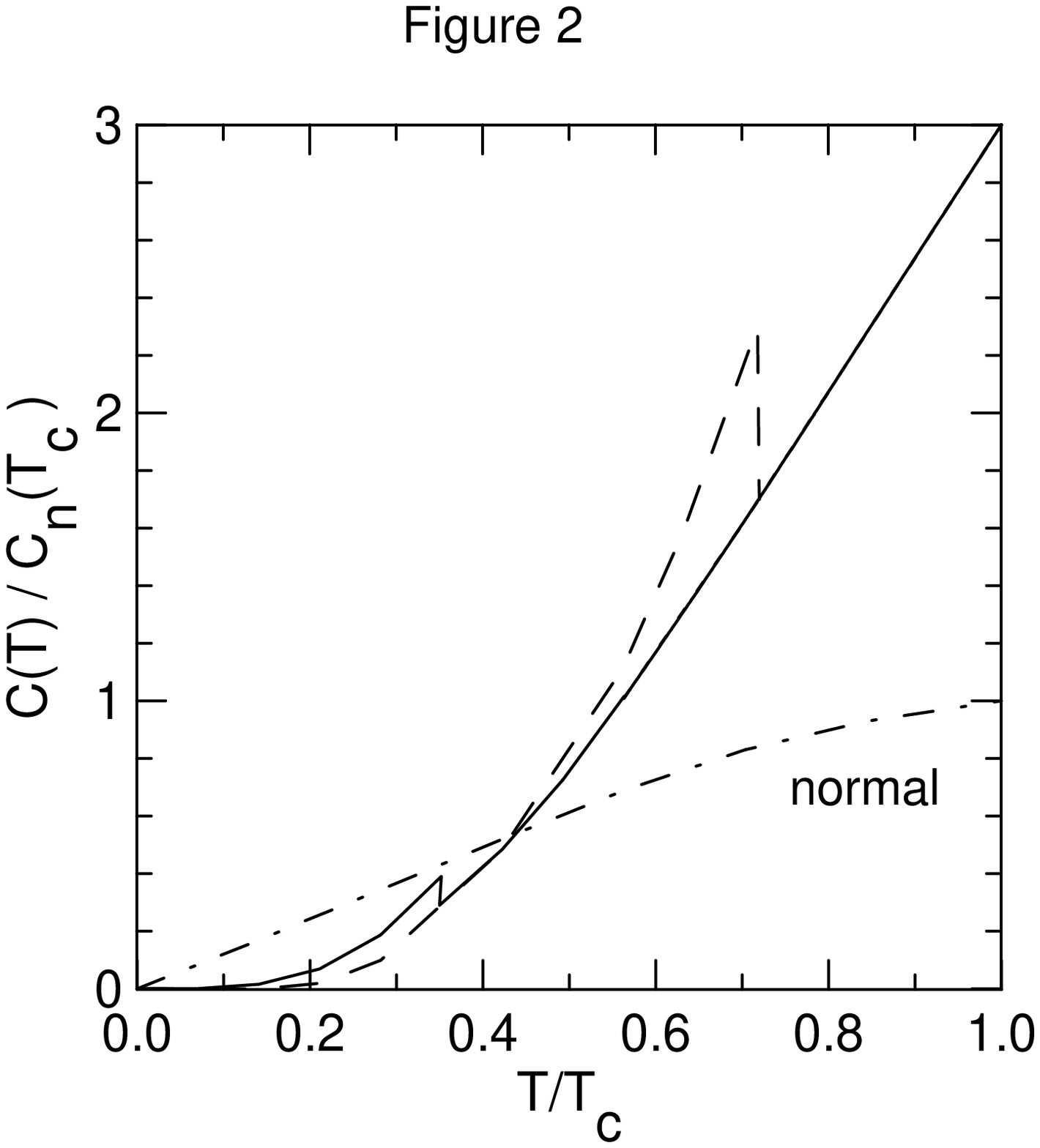}{1.0}    
\vskip -1.6cm

{ {\bf Fig. 2.}   Specific heat ratio $C(T)/C_n(T_c)$ versus $T/T_c$
for models 1(a)
and 1(b) on square lattice: 1(a)
(full line) and 
1(b)  (dashed line). The dashed-dotted line represents the 
result for the normal state 
   for comparison.}

\vskip 0.4cm

We solved  the coupled set of equations (\ref{131}) and (\ref{132})
numerically and
calculated the  gaps $\Delta_0$ and $\Delta_2$ at various temperatures for
$T<T_c$. 
We have performed  calculations    (1) on a perfect square lattice and 
(2)  in the presence of  an orthorhombic
distortion with Debye cut off 
$k_BT_D=0.02586 $ eV 
($T_D = 300$ K) in both cases. 
 The parameters for these two cases are the following:
(1) Square lattice $-$ (a)
$t=0.2586 $ eV, $\beta=1$,  $\gamma = 0$, $V_{0}=1.8t$,
and $V_2=0.73t$, $T_c = 71$ K, $T_{c1}$ = 25 K; 
(b) $t=0.2586 $ eV, $\beta=1$,  $\gamma = 0$, $V_{0}=1.92t$,
and $V_2=0.73t$, $T_c = 71$ K, $T_{c1}$ = 51 K; 
(2) Orthorombic distorsion $-$ (a)
$t=0.2586 $ eV, $\beta = 0.95$, and $\gamma=0$, $V_{0}=2.06t$,
and $V_2=0.97t$, $T_c$ = 70 K, $T_{c1} $ = 25 K;
(b)
$t=0.2586 $ eV, $\beta = 0.95$, and $\gamma =0$, $V_{0}=2.2t$,
and $V_2=0.97t$, $T_c$ = 70 K, $T_{c1} $ = 50 K.
 For a 
very weak $s$-wave ($d$-wave) coupling the only possible solution 
corresponds to $\Delta_0 =0$ ($\Delta_2 =0$).
We have 
studied the solution only when a coupling is allowed between 
Eqs. (\ref{131}) and (\ref{132}).

\vskip -2.5cm
\postscript{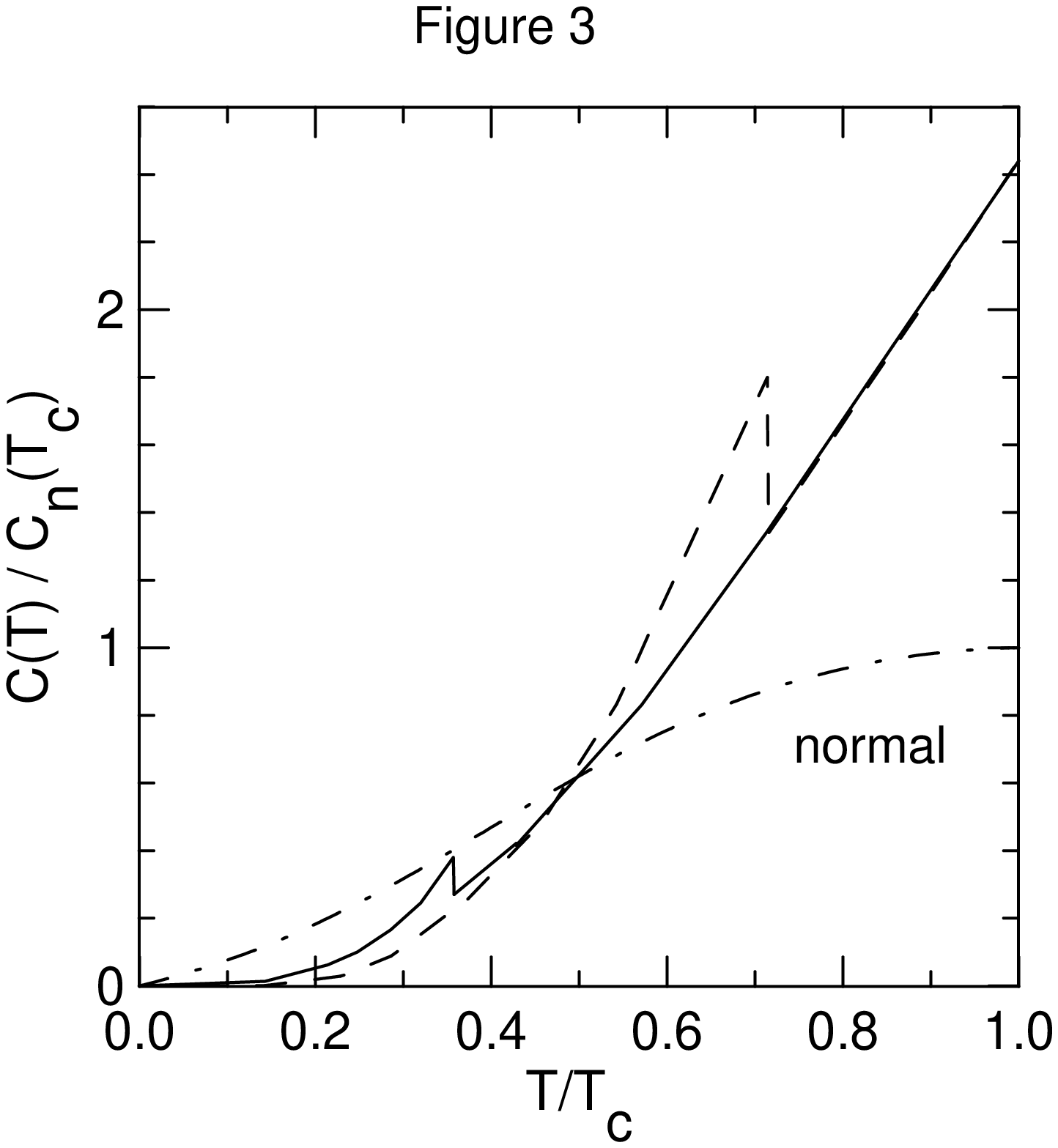}{1.0}    
\vskip -1.6cm

{ {\bf Fig. 3.}    3. Specific heat ratio $C(T)/C_n(T_c)$ versus $T/T_c$  for models 2(a)
and 2(b) for lattice with orthorhombic distortion:
for notations see Fig. 2.}

\vskip -2.5cm
\postscript{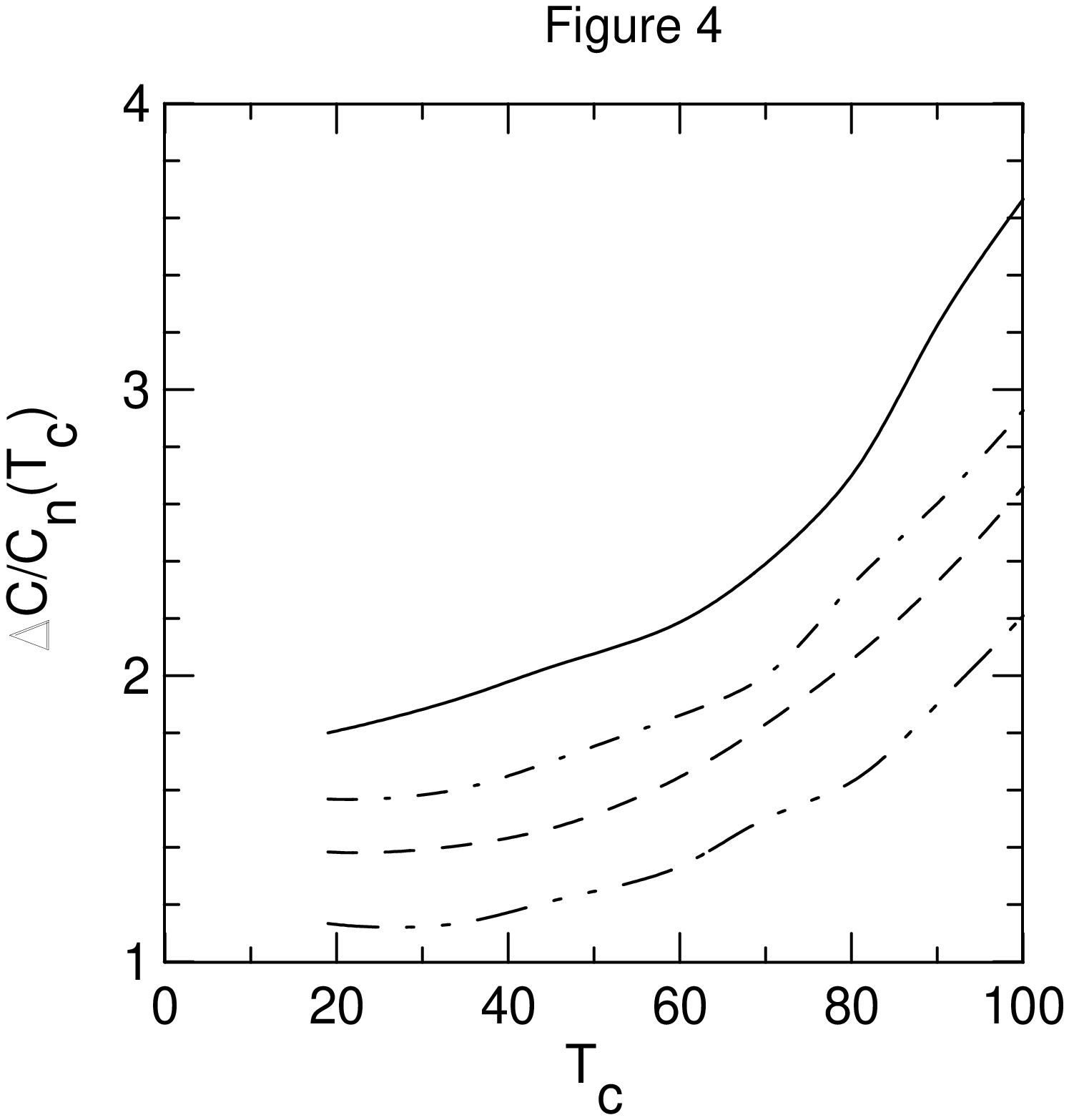}{1.0}    
\vskip -1.2cm

{ {\bf Fig. 4.} 4. Specific heat jump for different $T_c$ for pure $s$ and $d$ waves:
$s$ wave (solid line, square lattice),
$s$ wave (dashed line, orthorhombic distortion),
$d$ wave (dashed-dotted line, square lattice),
$d$ wave (dashed-double-dotted line, orthorhombic distortion).}

\vskip 0.4cm

The temperature dependencies of different $\Delta$'s have been studied in
details previously \cite{9a}. Here, for completeness, in Fig.  1 we plot the
temperature dependencies of different $\Delta$'s for the following two sets
of $s$-$d$ mixing corresponding to models 1(a) (full line) and 2(b) (dashed
line), respectively.  In the coupled  $ (d_{x^2-y^2}+is)$ wave
as temperature is lowered past $T_c$, the parameter 
$\Delta_2$ increases up to $T=T_{c1}.$ With further reduction of temperature,
with the appearance of the parameter $\Delta_0$, the $d$-wave component 
 begins to decrease and   the parameter $\Delta_2$ is suppressed in
the presence of a non-zero $\Delta_0$.

We also studied the effect of orthorhombicity in this problem. For fixed $V_0$
and $V_2$,  if a small orthorhombicity is introduced in the  model,  $T_c$
($T_{c1}$) decreases (increases). For example, if, in the model 1(a) above, we
vary  the parameter $\beta$ from 1 to 0.95, there is no coupled solution
involving $s$ and $d$ waves for $\beta <0.96$, where
 only the pure $s$-wave solution survives.  The
coupled solution is observed for $\beta \geq 0.97$ in this case.  For $\beta =
0.97$ (0.99), 
with the parameters of model 1(a), $T_c$ = 58 K (69 K), and $T_{c1}$ = 42 K
(28 K). So the effect of introducing the orthorhombicity in a model is to 
increase $T_{c1}$ and decrease $T_c$. As a consequence the ratio 
$T_{c1}/T_c$ increases.

In order to substantiate the claim of the second phase transition at
$T=T_{c1}$, we study the temperature dependence of specific heat in some
detail. The different superconducting and normal specific heats are plotted
in Figs. 2 and 3 for square lattice [models 1(a) and 1(b)] and orthorhombic
distortion [models 2(a) and 2(b)], respectively.  The superconducting
specific heat exhibits  an unexpected peculiar behavior.  In both cases the
specific heat exhibits two jumps $-$ one at $T_c$ and another at $T_{c1}$.
From  Eq. (\ref{sp}) and Fig.  1 we see that the temperature derivative of
$|\Delta_{\bf q}|^2$ has discontinuities at $T_c$ and $T_{c1}$ due to the
vanishing of $\Delta_2$ and $\Delta_0$, respectively, responsible for the two
jumps in specific heat.  For pure $d$ wave we find that the specific heat
exhibits a power-law dependence on temperature. However,  the exponent of
this dependence varies with temperature. For small $T$ the exponent is
approximately 2.5, and for large $T$ ($T\to T_c$) it is nearly 2.  In the
$(d+is)$-wave models, for $T_c > T > T_{c1}$ the specific heat exhibits
$d$-wave
power-law behavior. For $d$-wave models $C_s(T_c)/C_n(T_c)$ is a function of
$T_c$ and $\beta$. In Figs. 2  and 3  this ratio for the $d$-wave case, for
$T_c$ = 70 K, is approximately 3 (2.5) for $\beta =$ 1 (0.95).  In a
continuum calculation  this ratio was 2 in the absence of a van Hove
singularity \cite{c}. For $T<T_{c1}$, we find an exponential behavior in both
cases.

In Fig. 4 we study the jump $\Delta C$ in the specific heat at $T_c$ for pure
$s$- and $d$-wave superconductors as a function of $T_c$, where we plot
the ratio 
$\Delta C/C_n(T_c)$ versus $T_c$. For a BCS superconductor in the continuum
$\Delta C/C_n(T_c)$ = 1.43  (1.0) for $s$-wave ($d$-wave) superconductor
independent of $T_c$ \cite{c,t}. 
Because of the presence of the van Hove singularity in
the present model this ratio increases with $T_c$ as can be seen in Fig. 4.
For a fixed $T_c$, the ratio 
$\Delta C/C_n(T_c)$ is greater for square lattice ($\beta =1$) than 
that for a lattice with orthorhombic distortion ($\beta=0.95$) for both $s$ 
and $d$ waves. 
At $T_c$ = 100 K,  in the  $s$-wave ($d$-wave)
square lattice case this ratio could be as high as 3.63 (2.92). Many 
high-$T_c$ materials have produced a large value for this ratio.

\vskip -2.5cm
\postscript{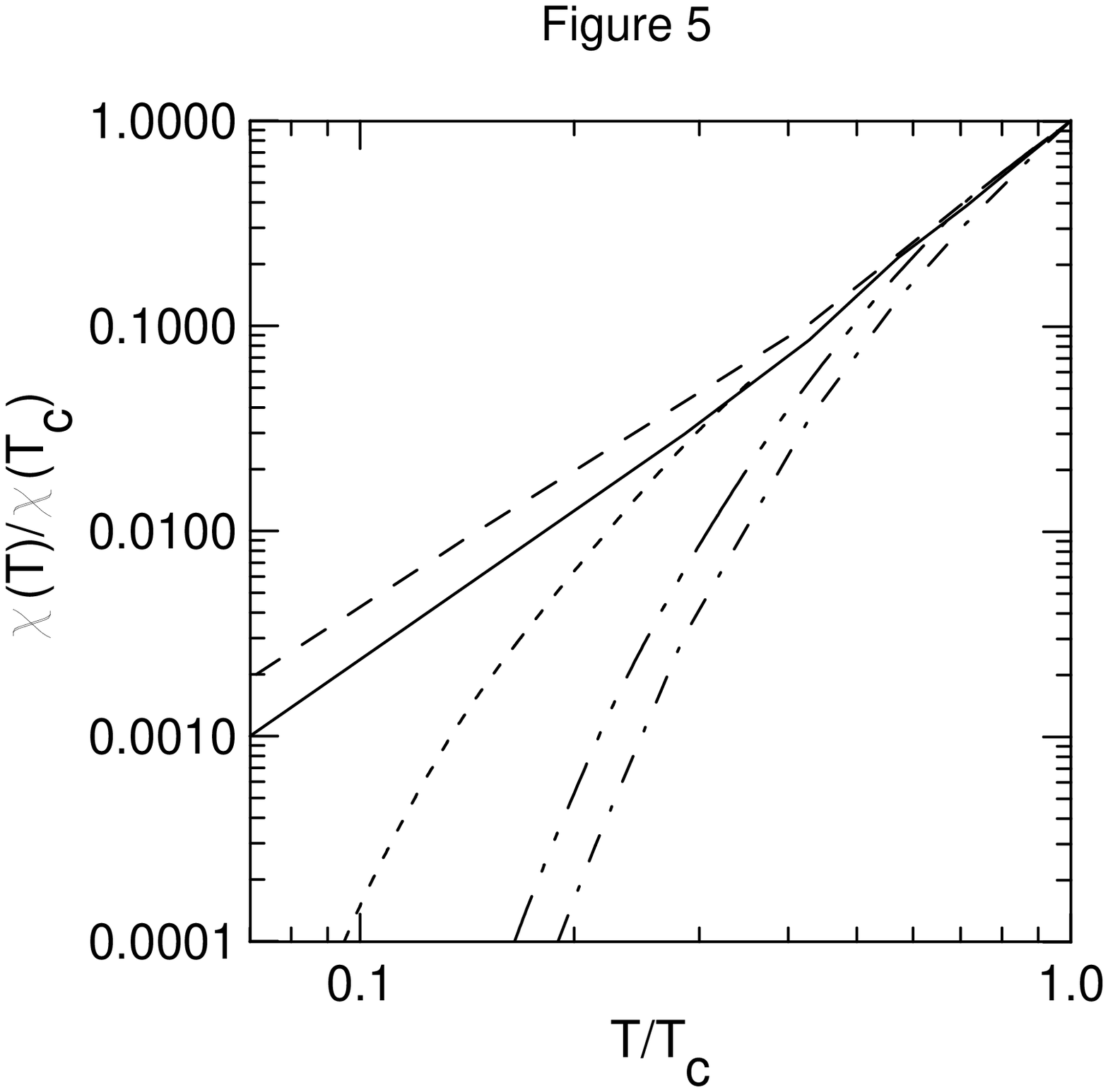}{1.0}    
\vskip -1.2cm

{ {\bf Fig. 5.}   Susceptibility  ratio $\chi(T)/\chi(T_c)$ versus $T/T_c$  
for pure $d$ wave (solid line, square lattice),
$d$ wave (dashed line, orthorhombic distortion),
pure $s$ wave (dashed-dotted line, square lattice),
$ (d_{x^2-y^2}+is)$ model 1(a) (dotted line),
 $ (d_{x^2-y^2}+is)$  model 2(b) (dashed-double-dotted line).
In all cases $T_c \approx 70$ K. }

\vskip 0.4cm

Next we study the temperature dependencies of spin susceptibility,
penetration depth, and thermal conductivity which we exhibit in Figs. 5 $-$ 7
where  we also plot the results for pure $s$ and $d$ waves   for
comparison.  In Figs. 5 $-$ 7, we show the results for pure $d$-wave 
cases on square lattice and with orthorhombic distortion, $(d+is)$ models 
1(a) and 2(b) mentioned above, and pure $s$-wave case on square lattice.
In all cases reported in these figures $T_c \approx 70$ K. 
For pure $d$-wave case  we obtained
power-law dependencies on temperature. The exponent for this power-law
scaling was independent of critical temperature $T_c$ but varied from a
square lattice to that with an orthorhombic distortion. In case of thermal
conductivity, the exponent for square lattice (orthorhombic distortion,
$\beta=0.95$)   is 2.2 (1.4).  For spin susceptibility, the exponent for square
lattice (orthorhombic distortion, $\beta$ = 0.95)   is 2.6 (2.4).  
For the mixed
$(d+is)$-wave case, $d$-wave-type power-law behavior is obtained for
$T_c>T>T_{c1}$ with the same exponent as in the pure $d$-wave case.  For
$T<T_{c1}$, there is no node in the present order parameter on the Fermi
surface and one has a typical  $s$-wave behavior.  A passage from
$d$-  to  $s$-type state at $T_{c1}$ represents an increase in order and
hence an increase in superconductivity \cite{c}.  As temperature decreases,
the system passes from the normal state to a $d$-wave state  at $T=T_c$ and
then to a $s$-wave-type state at $T=T_{c1}$ signaling a second phase
transition.

Until now no high-$T_c$ material has clearly shown two jumps in specific
heat  as noted in the present study. If this transition to a mixed state
occurs at a  low temperature this jump is expected to be  be small and
very high precision experimental data will be needed for its confirmation. 
The same will be needed for confirming this phase transition from a
measurement of Knight shift which should show a change from a power-law
to exponential dependence on temperature. However, if a transition from a
pure $d$-wave state to a mixed-symmetry state takes place in any
compound, they could be identified by a study of the temperature
dependencies of the observables considered in this work. 

\vskip -2.5cm
\postscript{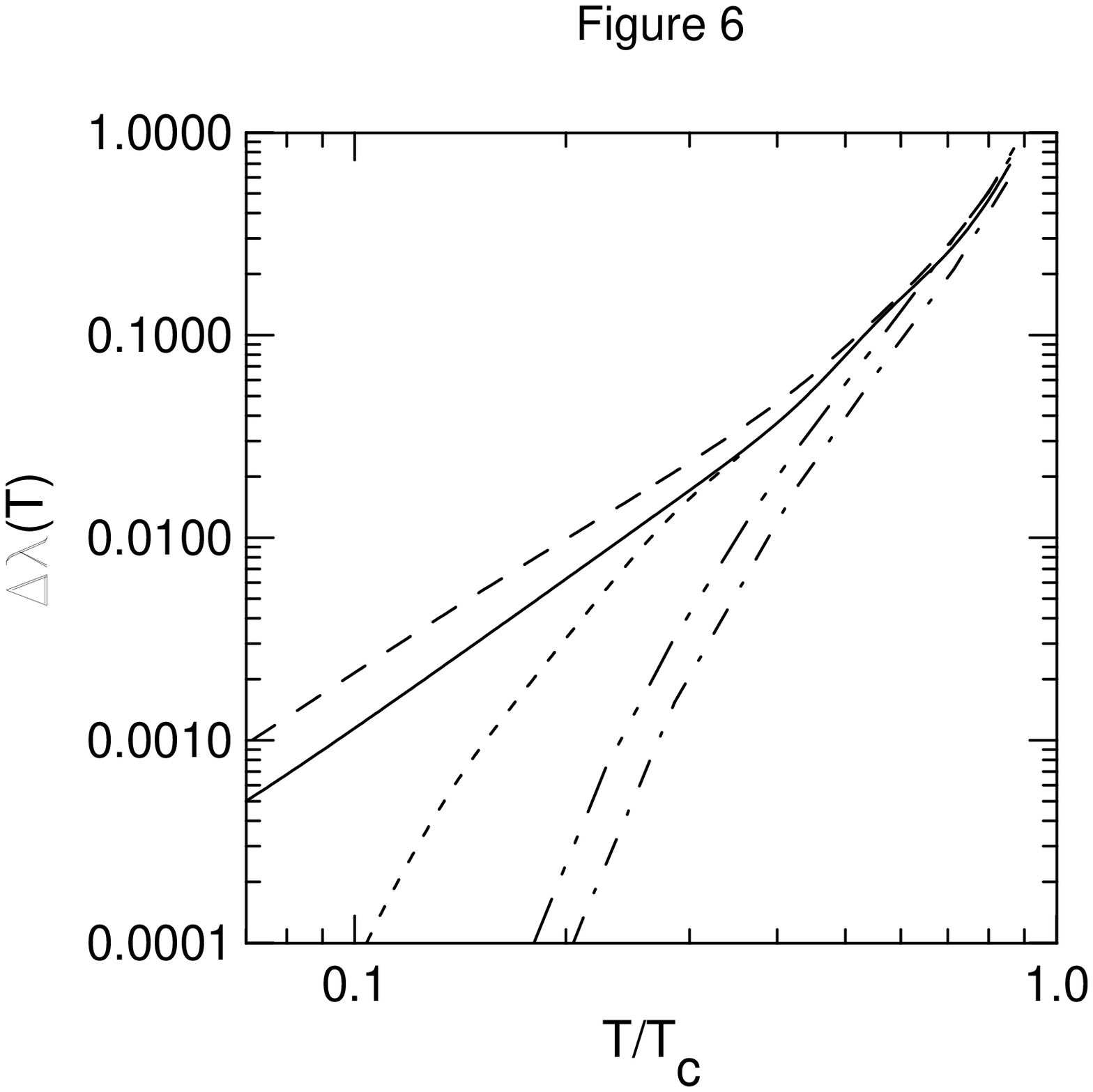}{1.0}    
\vskip -1.2cm

{ {\bf Fig. 6.}    Penetration depth  ratio $\Delta \lambda(T)\equiv
 [\lambda(T)-\lambda(0)]/\lambda(0)$ versus $T/T_c$  for different models.
For notation see Fig. 5.}

\vskip -2.5cm
\postscript{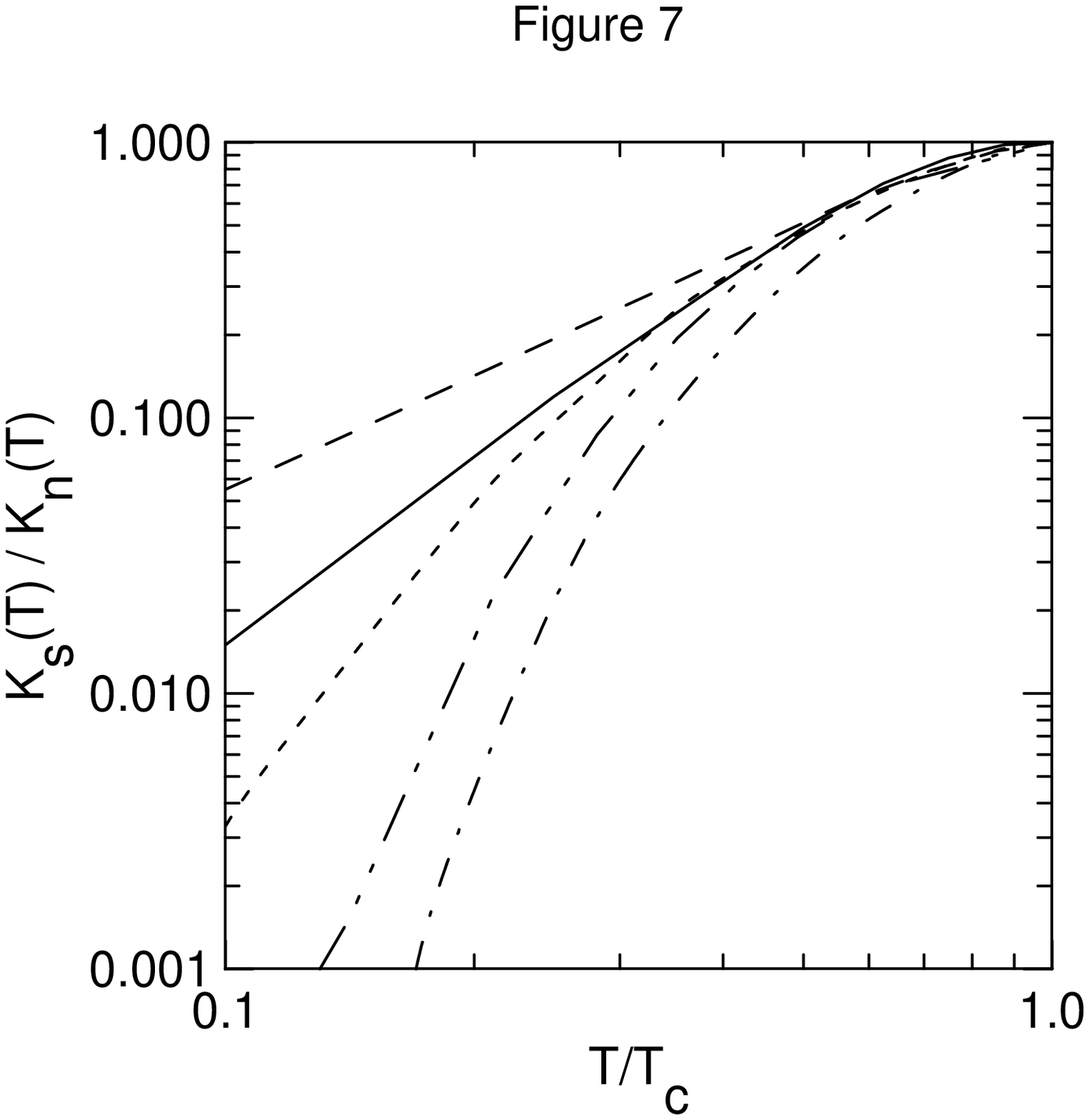}{1.0}    
\vskip -1.2cm

{ {\bf Fig. 7.}   Thermal conductivity   ratio $K_s(T)/K_n(T)$ versus $T/T_c$  
for different models.
For notation see Fig. 5.}

\vskip 0.4cm

In conclusion, we have studied the $ (d_{x^2-y^2}+is)$-wave
superconductivity employing
a two-dimensional tight binding BCS model on square lattice and also for 
orthorhombic distortion
and confirmed a  second second-order
phase transition at  $T=T_{c1}$ in the presence of a weaker $s$ wave.  We
have kept the $s$- and $d$-wave couplings in such a domain that a coupled
$ (d_{x^2-y^2}+is)$-wave solution is allowed. As temperature is lowered
past the first
critical temperature $T_c$, a weaker (less ordered) superconducting phase is
created  in  $ d_{x^2-y^2}$
 wave, which changes to a stronger (more ordered)
superconducting phase in   $ (d_{x^2-y^2}+is)$ wave at $T_{c1}$.     The
 $ (d_{x^2-y^2}+is)$-wave
state is similar to an   $s$-wave-type  state with no node in the order
parameter. The phase transition at $T_{c1}$ is also marked by power-law
(exponential) temperature dependencies of $ C(T), \chi(T)$, $\Delta
\lambda (T)$ and $K(T)$ for $T > T_{c1}$ ($<T_{c1} $).  
Furthermore, the effect of orthorhombic distortion is shown to increase 
the transition temperature between these two superconducting phases, thus
stabilizing the mixed state.

We thank 
  Conselho
Nacional de Desenvolvimento Cient\'{\i}fico e Tecnol\'ogico and Funda\c c\~ao
de Amparo \`a Pesquisa do Estado de S\~ao Paulo for financial support.

\end{document}